\documentstyle[12pt,psfig,epsf]{article}
\input epsf.tex

\begin{document}

\newcommand{\be}{\begin{equation}}
\newcommand{\ee}{\end{equation}}
\newcommand{\bea}{\begin{eqnarray}}
\newcommand{\eea}{\end{eqnarray}}
\def\href#1#2{#2} 

\baselineskip=15.5pt
\pagestyle{plain}
\setcounter{page}{1}

\begin{titlepage}

\begin{flushright}
PUPT-1835\\
DEMO-HEP 99/2 Jan.\\
hep-th/9902023
\end{flushright}
\vspace{14 mm}

\begin{center}
{\huge   Neumann boundary conditions from Born-Infeld dynamics}
\end{center}
\vspace{6 mm}
\begin{center}
{\large Konstantin G. Savvidy \footnote{ksavvidi@princeton.edu} }\\
\vspace{2mm}
{\it Joseph Henry Laboratories, Princeton University, Princeton,
New Jersey 08544}\\
\vspace{2mm}
{\large George K. Savvidy \footnote{savvidy@argo.nrcps.ariadne-t.gr}} \\
\vspace{2mm}
{\it National Research Center "Demokritos", Ag. Paraskevi, 15310 Athens, 
Greece}\\
\end{center}
\vspace{4 mm}
\begin{center}
{\large Abstract}
\end{center}
\noindent
We would like to show that certain excitations of the F-string/D3-brane 
system can be shown to obey   Neumann boundary conditions by considering 
the Born-Infeld dynamics of the F-string (viewed as a 3-brane cylindrically
wrapped on an $S_2$). In the paper by Callan and Maldacena it was shown that
excitations which are normal to both the string and the 3-brane behave as if
they had Dirichlet boundary conditions at the point of attachement. Here 
we show that excitations which are coming down the string with a 
polarization along a direction parallel to the brane are almost completely
reflected just as in the case of all-normal excitations, but the end of
the string moves freely on the 3-brane, thus realizing Polchinski's
open string    Neumann
boundary condition dynamically. In the low energy limit 
$ \omega \rightarrow 0$, i.e. for wavelengths much larger than the string
scale only a small fraction $ \sim \omega^4$ of the energy escapes in
the form of dipole radiation.
The physical interpretation is that a string attached to the 3-brane 
manifests itself as an electric charge, and waves on the string cause
the end point of the string to freely oscillate and therefore produce 
e.m. dipole radiation in the asymptotic outer region.

\end{titlepage}
\newpage

\section{ Setup}
Callan and Maldacena \cite{cm} showed that the Born-Infeld action,
when taken as the fundamental action, can be used to build a 
configuration with a semi-infinite 
fundamental string ending on a 3-brane\footnote{ We choose the 
favorite case of the 3-brane because, first of its non-singular
behaviour in SUGRA, and secondly it is suggestive of our own world
which is after all 3-dimensional.}, whereby the string is actually
made out of the brane wrapped on $S_2$ (see also \cite{gibbons}). 
The relevant action can be
obtained by computing a simple Born-Infeld determinant, dimensionally
reduced from 10 dimensions
\be
L=-{1\over(2\pi)^3g_s} \int d^4 x \sqrt{1- \vec{E}^2 + (\vec{\partial}x_9)^2},
\ee
where $g_s$ is the string coupling ($\alpha^{\prime} =1$).

The above mentioned theory contains 6 scalars $ x_4,....,x_9~,$ which are 
essentially Kaluza-Klein remnants from the 10-dimensional $N=1$ 
electrodynamics after dimensional reduction to $3+1$ dimensions. As is
well known these extra components of the e.m. field $A_4,...,A_9$ describe
the transverse deviations of the brane $x_4,...,x_9$.

The solution, which satisfies the BPS conditions, is necessarily also
a solution of the linear theory, the $N=4$ {\it super-}Electrodynamics,
\be
\vec{E} = \frac{c}{r^2}\vec{e_r}~,~~~~\vec{\partial}x_9 
= \frac{c}{r^2}\vec{e_r}~,~~~~x_9 = -\frac{c}{r}~,
\label{spike}
\ee
where $c=\pi g_s$ sets the distance scale, and in what follows we
drop it. 
Here the scalar field represents the geometrical spike,
and the electric field insures that the string carries uniform NS 
charge along it. The RR charge of the 3-brane cancels out on the 
string, or rather the tube behaves as a kind of RR dipole whose 
magnitude can be ignored when the tube becomes thin. Also, it is
seen in \cite{cm} that the infinite electrostatic energy of the point
charge can be reinterpreted as being due to the infinite length of the
attached string. The energy per unit length comes from the electric field
and corresponds exactly to the fundamental string tension.

Polchinski, when he introduced D-branes as objects on which strings can
end, required that the string have Dirichlet (fixed) boundary conditions
for coordinates normal to the brane, and   Neumann (free) boundary 
conditions for coordinate directions parallel to the brane \cite{rr,pol,leigh}. 
It was
shown in \cite{cm} that small fluctuations which are normal to both the 
string and the brane are mostly reflected back with a 
$phase~ shift  \rightarrow \pi$
which indeed corresponds to Dirichlet boundary condition.
See also \cite{larus} and \cite{rey} for 
a supergravity treatment of this problem.

In this paper we will show that P-wave excitations which are coming down 
the string with a
polarization along a direction parallel to the brane are almost completely
reflected just as in the case of all-normal excitations, but the end of
the string moves freely on the 3-brane, thus realizing Polchinski's
open string    Neumann boundary condition dynamically. As we will see  
a superposition of excitations of the 
scalar $x_9$ and of the e.m. field reproduces the required behaviour,
e.g. reflection of the geometrical fluctuation with a 
$phase~ shift  \rightarrow 0$
(Neumann boundary condition)\footnote{ 
This problem was also considered in \cite{rey} where the e.m. field is 
integrated out to produce an effective lagrangian for the scalar field
only. The other essensial difference with us is that 
we consider $P$-wave modes of the scalar field which describe physical 
transverse fluctuations of the F-string and not the $S$-wave modes  
which do not correspond to physical excitations of the string. }.

In addition we observe e.m. dipole radiation which escapes to infinity
from the place where the string is attached to the 3-brane. 
We shall see that in the low energy limit
$ \omega \rightarrow 0$, i.e. for wavelengths much larger than the string
scale a small fraction $ \sim \omega^4$ of the energy escapes to infinity in
the form of e.m. dipole radiation.
The physical interpretation is that a string attached to the 3-brane
manifests itself as an electric charge, and waves on the string cause
the end point of the string to freely oscillate and therefore produce
e.m. dipole radiation in the asymptotic outer region of the 3-brane.
Thus not only in the static case, but also in a more general dynamical 
situation the above interpretation remains valid.
This result provides additional support to the idea that 
the electron may be understood as the end of a fundamental string 
ending on a D-brane.

\section{The Lagrangian and the equations}

Let us write out the full Lagrangian which contains both 
electric and magnetic fields, plus the scalar $x_9 \equiv \phi$
\bea
\nonumber
L=-\int d^4x \sqrt{Det}~,~~~~~~~~~~~~~~~~~~~~~~~~~~~~~
~~~~~~~~~~~~~~~~~~~~~~~~~~~~~~~~~~\\
\nonumber
where~~~Det=1+ \vec{B}^2 - \vec{E}^2 -(\vec{E} \cdot \vec{B})^2 -
   (\partial_0 \phi)^2(1+\vec{B}^2) +(\vec{\partial}\phi)^2\\
   +(\vec{B}\cdot \vec{\partial} \phi)^2 - (\vec{E}\times \vec{\partial}\phi)^2+
   2\partial_0 \phi(\vec{B}[\vec{\partial} \phi\times\vec{E}]) 
\label{lagrfull}
\eea
We will proceed by adding a fluctuation to the background values 
(\ref{spike}) : 
$$ \vec{E}=\vec{E}_0 + \delta \vec{E},~~\vec{B}= 
\delta \vec{B},~~\phi=\phi_0 +\eta~.$$
Then keeping only terms in the $Det$  which are linear and quadratic
in the fluctuation we will get 
\bea
\delta Det =  \delta\vec{B}^2 - \delta\vec{E}^2 - 
(\vec{E_{0}} \delta\vec{B})^2  - (\partial_0 \eta)^2 +   
(\vec{\partial}\eta)^2 \\ \nonumber
+ (\delta\vec{B} \vec{\partial} \phi)^2
- (\vec{E_0} \times \vec{\partial}\eta)^2 - 
(\delta\vec{E} \times \vec{\partial} \phi)^2 
-2(\vec{E_0} \times \vec{\partial}\eta)
(\delta\vec{E} \times \vec{\partial}\phi)\\ \nonumber
-2(\vec{E_0} \delta\vec{E}) + 2 (\vec{\partial}\phi \vec{\partial}\eta)
\eea
Note that one should keep the last two linear terms because they produce 
additional quadratic terms after taking the square root.
These terms involve the longitudinal polarization of the 
e.m. field and cancel out at quadratic order.
The resulting quadratic Lagrangian is 
\be
2L_q = \delta\vec{E}^2(1+(\vec{\partial} \phi)^2) - \delta\vec{B}^2 + 
(\partial_0\eta)^2 -
(\vec{\partial}\eta)^2(1-\vec{E_0}^2) + \vec{E_0}^2 
( \vec{\partial}\eta \cdot \delta\vec{E})~~.
\label{full}
\ee
Let us introduce the gauge potential for the fluctuation part of the e.m.
field as $(A_0,\vec{A})$ and substitute the values of the background 
fields from (\ref{spike})
\be
2L_q = (\partial_0 \vec{A}- \vec{\partial}A_0)^2 (1+{1\over r^4})-
       (\vec{\nabla}\times\vec{A})^2 + (\partial_0\eta)^2 -
       (\vec{\partial}\eta)^2(1-{1\over r^4}) + 
       {1\over r^4} (\partial_0 \vec{A}- \vec{\partial}A_0)
       \cdot \vec{\partial}\eta ~~.
\label{fluctlag}
\ee
The equations that follow from this lagrangian contain dynamical 
equations for the vector potential and for the scalar field, 
and a separate equation which represents a constraint. These
equations in the Lorenz gauge
$\vec{\partial}~\vec{A}=\partial_0 A_0$ are
\bea
\nonumber
~~-\partial_0^2 \vec{A}(1+{1\over r^4}) + \Delta\vec{A}~+~~~
{1\over r^4}\vec{\partial}\partial_0(A_0+\eta) = 0~~~~~&{(\bf\alpha )} \\
\nonumber
~~-\partial_0^2 A_0 + \Delta A_0 + 
\vec{\partial} {1\over r^4} \vec{\partial}(A_0+\eta) - 
\vec{\partial} {1\over r^4} \partial_0\vec{A} = 0~~~~~&{(\bf\beta )} \\ 
\nonumber
~~-\partial_0^2 \eta~ + \Delta \eta~~ - 
\vec{\partial} {1\over r^4} \vec{\partial}(A_0+\eta) + 
\vec{\partial} {1\over r^4} \partial_0\vec{A} = 0~~~~~&{(\bf\gamma )}
\eea
Equation $(\beta)$ is a constraint: the time derivative of
the {\it lhs} is zero, as can be shown using the equation of motion 
$(\alpha)$.

Let us choose $A_0=-\eta$. This condition can be viewed as (an attempt to)
preserve the BPS relation which holds for the background: 
$\vec{E}=\vec{\partial } \phi$. Another point of view is that this fixes
the general coordinate invariance which is inherent in the Born-Infeld 
lagrangian in such a way as to make the given perturbation to be
normal to the surface. Of course transversality is insured automatically
but this choice makes it explicite. The general treatment of this 
subject can be found in \cite{sch}.

With this condition the equations $(\beta)$ and $(\gamma)$ become the same, and
the first equation is also simplified:
\bea
-\partial_0^2 \vec{A}(1+{1\over r^4}) + ~\Delta\vec{A} = 0~, \\
-\partial_0^2 \eta + \Delta \eta +
\vec{\partial} {1\over r^4} \partial_0\vec{A} = 0~.
\label{seqns}
\eea
This should be understood to imply that once we obtain a solution,
$A_0$ is determined from $\eta$, but in addition we are now obliged to 
respect the gauge condition which goes over to 
$\vec{\partial}\vec{A}=-\partial_0 \eta$.

\section{Neumann boundary conditions and dipole radiation}
We will seek for a solution with definite energy 
(frequency $w$) in the following form: $\vec{A}$ should have
only one component $A_z$, and $\eta$ be an $l=1$ spherical $P$-wave
$$
A_z= \zeta(r)~e^{-i\omega t}~~,~~~\eta={z\over r}~\psi(r) ~e^{-i\omega t}
$$
The geometrical meaning of such a choice for $\eta$ is explained in Fig 1.
With this ansatz the equations become
\bea
(1+{1\over r^4})~\omega^2 \zeta + 
{1\over r^2}\partial_r(r^2 \partial_r \zeta)=0~~\\
{z\over r}~ \omega^2 \psi + 
{z\over r}~ {1\over r^2}\partial_r(r^2 \partial_r \psi) +
{z\over r}~ {2\over r^2}~\psi - i \omega \partial_z({\zeta\over r^4}) = 0~,
\eea
with the gauge condition becoming $\partial_r\zeta =i \omega \psi$.
It can be seen again, that the second equation follows from the first by
differentiation. This is because the former coincides with the constraint 
in our anzatz.

\begin{figure}
\centerline{\hbox{\psfig{figure=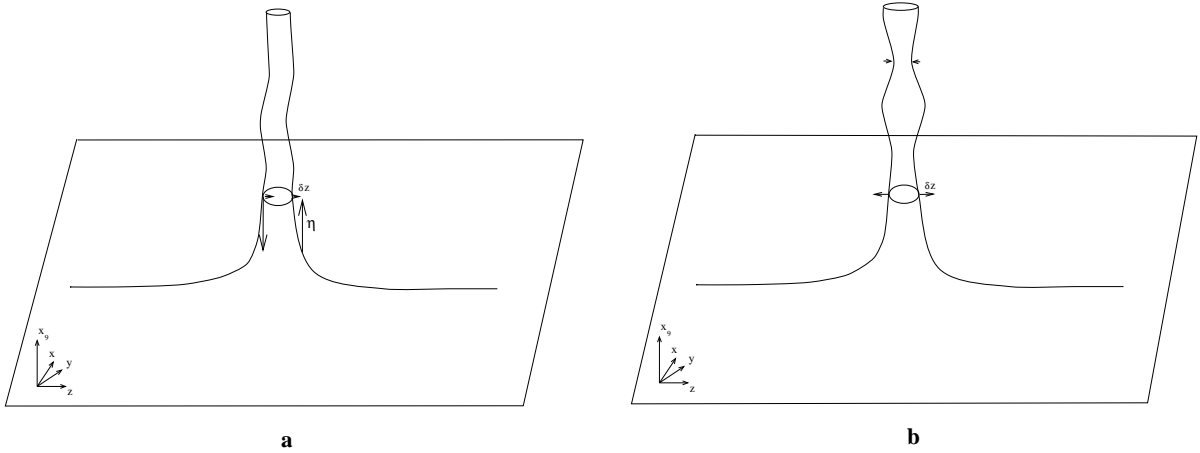,height=6cm,angle=0}}}
\caption[fig1]{ In order that all points on the $S_2$ section of the
tube ( which is schematically shown here as a circle) move all in
the same direction $\hat z$ by an equal distance $\delta z$, the field 
$\eta$ has to take on different values at, say, opposite points of the 
$S_2$. In effect, 
$ \eta=\delta{-1\over r}={1 \over r^2}{z \over r}\delta z$. 
If, on 
the other hand, it is taken to be an $S$-wave (as in the paper \cite{rey})
that would correspond to 
Fig 1b, which at best is a problematic `internal' degree of freedom of the
tube.
} 
\label{fig1}
\end{figure}

Therefore  the problem is reduced to finding the solution of a 
single scalar equation, and determining the other fields through
subsidiary conditions. The equation itself surprisingly turned out to be 
the one familiar  from \cite{cm} for the transverse fluctuations.
There it was solved by going over to a coordinate $\xi$ which
measures the distance radially along the surface of the brane
$$
\xi(r) = \omega \int\limits_1^r du \sqrt{1 +{1 \over u^4}}~~,
$$
and a new wavefunction
$$
\tilde{\zeta}=\zeta~(1+ r^4)^{1/4}~.
$$
This coordinate behaves as $\xi \sim r$ in the outer region 
($r\rightarrow\infty$)
and $\xi \sim -1/r$ on the string ($r\rightarrow 0$). 
The exact symmetry of the equation
$r \leftrightarrow 1/r$ goes over to $ \xi \leftrightarrow - \xi$. 
The equation, when written in this coordinate becomes just the free 
wave equation, plus a narrow symmetric potential at $\xi \sim 0$
\be
-{d^2\over d\xi^2}\tilde{\zeta}+\frac{5/\omega^2}{(r^2+1/r^2)^3}\tilde{\zeta}=0~.
\ee
The asymptotic wave functions can be constructed as plain waves in $\xi$, 
$$
\zeta(r) = (1 + r^4)^{-1/4} e^{\pm i \xi(r)}~,
$$
or in the various limits:
\bea
\nonumber
r\rightarrow ~0~~~~\zeta \sim ~~e^{\pm i \xi(r)}~,\\
\nonumber
r\rightarrow \infty~~~~\zeta \sim {1 \over r}e^{\pm i \xi(r)}~.
\eea
These formulae give us the asymptotic wave function in the regions 
$ \xi \rightarrow \pm \infty$, 
while around $\xi = 0~(r=1)$ there is a symmetric repulsive potential
which drops very fast $ \sim 1/\xi^6$ on either side of the origin.
The scattering is described by a single dimensionless parameter 
$\omega \sqrt{c}$, and in the limit of small $\omega$ and/or coupling 
$c=\pi g_s$ the potential becomes narrow and high, and  can be 
replaced by a $\delta$-function with an equivalent  area 
$\sim {1\over \omega \sqrt{g_s}}$ under the curve.
Therefore the scattering matrix becomes almost
independent of the exact form of the potential. The end result is that
most of the amplitude is reflected back with a phase shift close to $\pi$, thus
dynamically realizing the Dirichlet boundary condition in the low energy
limit. 

In order to obtain $\psi$ (and $\eta$) we need to differentiate
$\zeta$ with respect to $r$:
\be
i\omega\psi= {-1\over4} {4r^3\over(1+r^4)^{5/4}}e^{\pm i \xi(r)}
\pm{i\omega\over(1+r^4)^{1/4}}
(1 +{1 \over r^4})^{1/2} e^{\pm i \xi(r)}~~.
\ee
Again it is easy to obtain the simplified limiting form:
\bea
\nonumber
r\rightarrow 0~~~~i \omega \psi  
\sim (-r^3 \pm {i\omega \over r^2}) e^{\pm i \xi(r)}\\
\nonumber
r\rightarrow \infty~~~~i \omega \psi 
\sim ({-1 \over r^2} \pm {i\omega \over r}) e^{\pm i \xi(r)}
\eea 
This brings about several consequences for 
$\psi$. Firstly, it causes $\psi$ to grow as $\sim 1/r^2$ as 
$r\rightarrow 0$. This is the correct behaviour because when  converted to
displacement in the $z$ direction, it means constant amplitude.
Secondly, the $i$ that enters causes the superposition of the 
incoming and reflected waves to become a cosine from a sine, as is
the case for $\zeta$ waves. This corresponds to a $0$ phase shift and 
implies   Neumann boundary condition for the $\eta$ wave (Fig 2).

Because of the $\omega$ factor in the gauge condition we 
need to be careful about normalizations, thus we shall choose to 
fix the amplitude of the $\eta$ wave to be independent of $\omega$. 
Then the magnitude of the e.m. field in the inner region becomes 
independent of $\omega$ as well. 
Combined with the transmission factor, proportional to 
$\omega\sqrt{c}$, this gives the correct dependence of the total power emitted
by an oscillating charge $\sim \omega^4 g_s^2$.

\begin{figure}
\centerline{\hbox{\psfig{figure=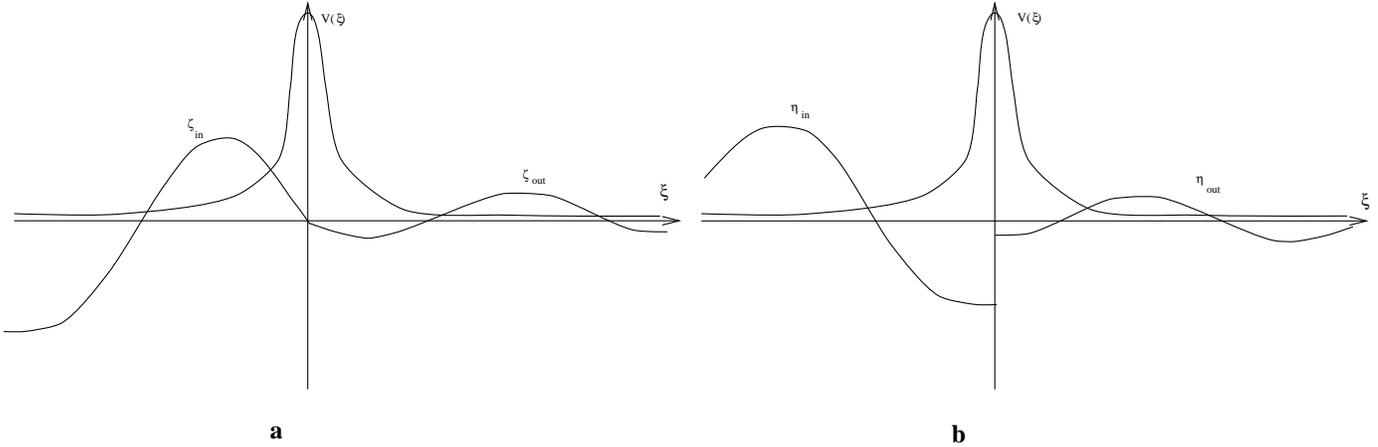,height=6cm,angle=0}}}
\caption[fig2]{ The figure 2a depicts the scattering of the 
$\zeta$ wave. Note the discontinuity in the derivative which is 
proportional to $1/\omega$. Figure 2b shows the scattering of
the $\eta$ wave. Being the derivative of $\zeta$, it 
results in a discontinuity of the function itself,
making it into a cosine, which means free (Neumann) 
boundary condition at $\xi =0$.
} 
\label{fig2}
\end{figure}

In conclusion, we need to analyze the outgoing scalar wave. This wave has 
both real and imaginary parts, the former is from differentiating the phase,
while the latter is from the prefactor. The imaginary part is $\sim 1/r^2$
which drops faster than radiation. The real part 
does contribute to the radiation at spatial infinity,
as can be shown from the integral of the energy density
$\int (\partial_r \eta)^2 dr \sim \int \omega^4/r^2 \cdot 4\pi r^2 dr$.
This is not altogether surprising, as we are dealing with a
supersymmetric theory where the different fields are tied together.
Thus the observer at spatial infinity will see both an
electromagnetic dipole radiation field and a scalar wave.


The problem of longitudinal fluctuations was treated in \cite{rey}, though 
not in a completely satisfactory way. There the scalar field was taken to be
an $S$-wave, which as should be apparent from Fig 1b, does not correspond 
to the string oscillating as a whole. In addition the electromagnetic field
was effectively integrated out, thus one cannot obtain the dipole 
radiation at spatial infinity.

\section{Acknowledgements}
KS was partially supported by National Science Foundation grant PHY98-02484.
One of us (KS) would like to thank C. Callan for extensive discussions.
We also thank G. Gibbons, I. Klebanov, S. Lee and L. Thorlacius 
for useful discussions.

\end{document}